\documentclass[a4paper,11pt]{article}
\usepackage{pos}

\title{Overview on Efforts for a Second Detector at the Electron-Ion Collider (EIC)}

\author*[a]{Jihee Kim}

\affiliation[a]{Department of Physics, Brookhaven National Laboratory, Upton, NY 11973, U.S.A.}

\emailAdd{jkim11@bnl.gov}

\abstract{The Electron-Ion Collider (EIC) will provide a unique experimental platform to explore the properties of gluons in nucleons and nuclei, offering new insights into their structure and dynamics. The EIC community has outlined a detailed physics program and the demanding detector requirements in a comprehensive detailed document. The primary general-purpose detector, ePIC, is designed to support a broad range of physics studies. However, there is strong community support for a second detector at the EIC to further enhance the scientific capabilities of the facility. A second detector would provide cross-checks and systematic controls for potential discoveries, while incorporating complementary technologies to address physics measurements that may be underrepresented by ePIC. In particular, it would improve forward detector acceptance at low transverse momentum ($p_T$) and enable more precise measurements in exclusive, diffractive, and tagging physics. This talk will provide a general overview of the second detector and outline its potential capabilities, highlighting key areas of the physics program it could enhance.}

\FullConference{
}

\begin{document}
\maketitle

\section{Introduction}
The Electron-Ion Collider (EIC) was identified as the highest-priority new construction project in the 2015 U.S. Nuclear Physics Long Range Plan~\cite{osti_1296778}, and its scientific significance was strongly endorsed by a 2018 review from the National Academy of Sciences~\cite{NAP25171}. The EIC will serve as a high-energy, high-luminosity experimental facility aimed at probing the properties of gluons within nucleons and nuclei, thereby advancing our understanding of their structure and dynamics. The scientific goals of the EIC were detailed in a community-driven White Paper~\cite{Accardi:1498519}, while comprehensive detector requirements and possible technologies were outlined in a Yellow Report~\cite{ABDULKHALEK2022122447}. The resulting general-purpose detector, ePIC, is designed to support a broad range of physics investigations and is planned for deployment at a single interaction point, IP-6. Although the EIC facility can accommodate two interaction regions (IR-6 and IR-8) and corresponding interaction points (IP-6 and IP-8), the current project scope includes only one detector.

\section{Motivation for a 2nd EIC Detector}
The broader EIC community strongly advocates the construction of a second detector, as detailed in a dedicated chapter of the Yellow Report~\cite{ABDULKHALEK2022122447}. There are several advantages to having two general-purpose collider detectors. First, independent measurements from two detectors help prevent erroneous conclusions caused by analysis errors, instrumental malfunctions, or statistical fluctuations—especially when investigating new phenomena that require cross-validation. Second, complementary detector designs enable cross-calibration, thereby reducing systematic uncertainties. Improvements in precision arise not only from increased data statistics but also from independent measurements that help identify and control detector-specific biases. Third, a second detector can be optimized for a distinct set of physics goals, thereby expanding the overall scientific reach of the EIC. Complementary programs and unique scientific motivations further strengthen the case for funding a second interaction region and detector. Additionally, deploying different technologies across two detectors mitigates risk and enhances the overall robustness of the EIC program.

\section{Complementarity of Technologies}
As previously discussed, complementarity between the two EIC detectors can be achieved by employing different detector technologies. Below are several illustrative examples, highlighting only a subset of the 36 subsystems included in the ePIC detector design:

\begin{description}
\item[Magnet:] While ePIC utilizes a 1.7~T solenoid (with peak fields approaching approximately 2~T), the second EIC detector could feature a larger solenoid with a magnetic field strength of 2–3~T and an increased radius. This design would improve tracking precision, provide additional space for detector services, and allow for greater detector depth. However, it also introduces technical challenges and risks, including the complexity of manufacturing a larger-radius, higher-field solenoid.

\item[Tracker:] A complementary tracking system might employ alternative technologies, such as a gaseous detector—e.g., a time projection chamber (TPC) or drift chamber—combined with outer layers of silicon-based precision trackers. This configuration could enhance pattern recognition, improve tracking efficiency, and enable particle identification of low-momentum particles through $dE/dx$ measurements.

\item[Particle Identification (PID):] PID systems could be optimized for different momentum regimes relative to ePIC. For instance, the second detector might emphasize high-momentum PID in the forward region using a simplified Ring Imaging Cherenkov (RICH) detector. Additionally, a stronger focus on time-of-flight (TOF) technology—with an ambitious target time resolution of around 10 picoseconds—could substantially improve PID performance in the low-intermediate momentum range.

\item[Barrel Hadronic Calorimeter (HCAL):] Instead of a conventional hadronic calorimeter, the second detector might consider integrating a dedicated muon detection system in the barrel region. This would enhance muon identification capabilities and provide additional sensitivity in physics channels involving heavy flavor or electroweak final states.
\end{description}

This complementarity in detector technologies—despite the detectors pursuing similar experimental objectives—plays a crucial role in cross-calibration. It helps to reduce systematic uncertainties associated with relying on a single detector configuration, especially when combining data from both the first and second EIC detectors for the same measurements.

\section{Complementarity of Interaction Regions}
The pre-conceptual design of IR-8 specifically improves acceptance for low transverse momentum protons and light nuclei in exclusive reactions at very low momentum transfer ($t$), as well as for nuclear breakup products in incoherent diffractive processes. This enhancement significantly increases the potential for detailed physics studies and tagging measurements at the EIC~\cite{ActaPhysPolBProcSuppl.18.1-A1}.

\begin{figure}[b]
    \centering
    \includegraphics[width=0.5\linewidth]{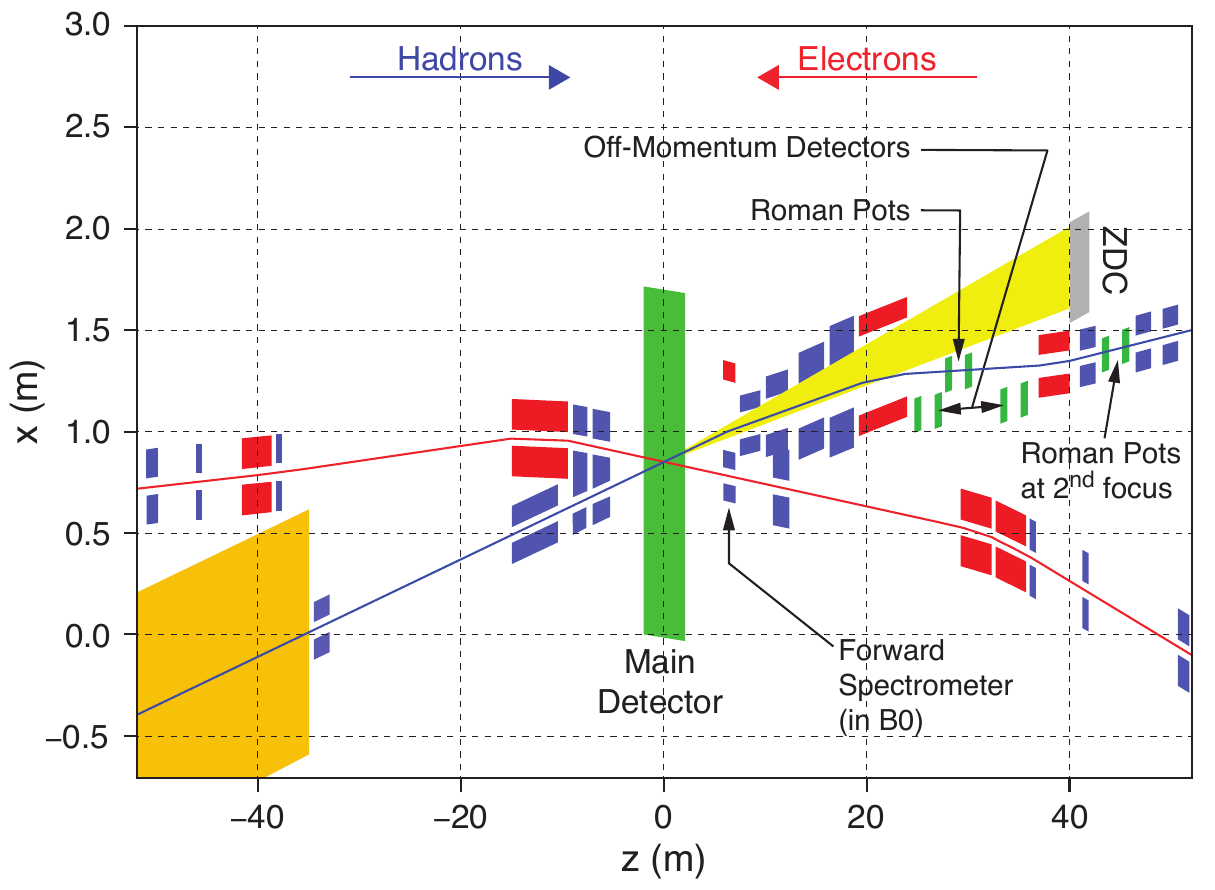}
    \caption{Schematic layout of the second interaction region (IR-8), designed with a 35~mrad crossing angle. The diagram illustrates the accelerator lattice, potential detector locations, and the position of the secondary beam focus along the hadron beamline. Adapted from~\cite{PhysRevD.111.072013}}
    \label{fig:ir8}
\end{figure}

The interaction regions IR-6 and IR-8 share luminosity between their respective detectors at the same center-of-mass energy and present comparable advantages and challenges from an accelerator design perspective. However, their configurations differ in key design features. IR-6, with a 25~mrad crossing angle, allows transverse momentum ($p_T$) measurements in the range of 0.2–1.3~GeV and provides neutron acceptance up to a polar angle of approximately $\theta \sim 4.5^\circ$.

In contrast, IR-8, illustrated in Figure~\ref{fig:ir8}, is constrained by the geometry of the existing experimental hall and tunnel, necessitating a larger optimal crossing angle of 35~mrad. This increased angle introduces distinct blind spots in pseudorapidity and makes achieving acceptance at high pseudorapidity values (e.g., $\eta \sim 3.5$) in the central detector more challenging. To compensate, the proposed hadron beamline for IR-8 incorporates an optical configuration with a secondary focus approximately 45~m downstream of IP-8, created by additional dipole and quadrupole magnets within the lattice. This design feature enhances forward detection acceptance despite the larger crossing angle.

The concept of a secondary focus is to create a narrow beam profile in the transverse plane—similar to the beam focus at the interaction point—enabling the detection of particles scattered at very small angles (near $\sim 0$~mrad) with minimal changes in magnetic rigidity. This feature provides the second EIC detector with complementary capabilities to the first, particularly benefiting exclusive, tagging, and diffractive physics programs.

\begin{figure*}[b]
    \centering
    \includegraphics[width=0.9\linewidth]{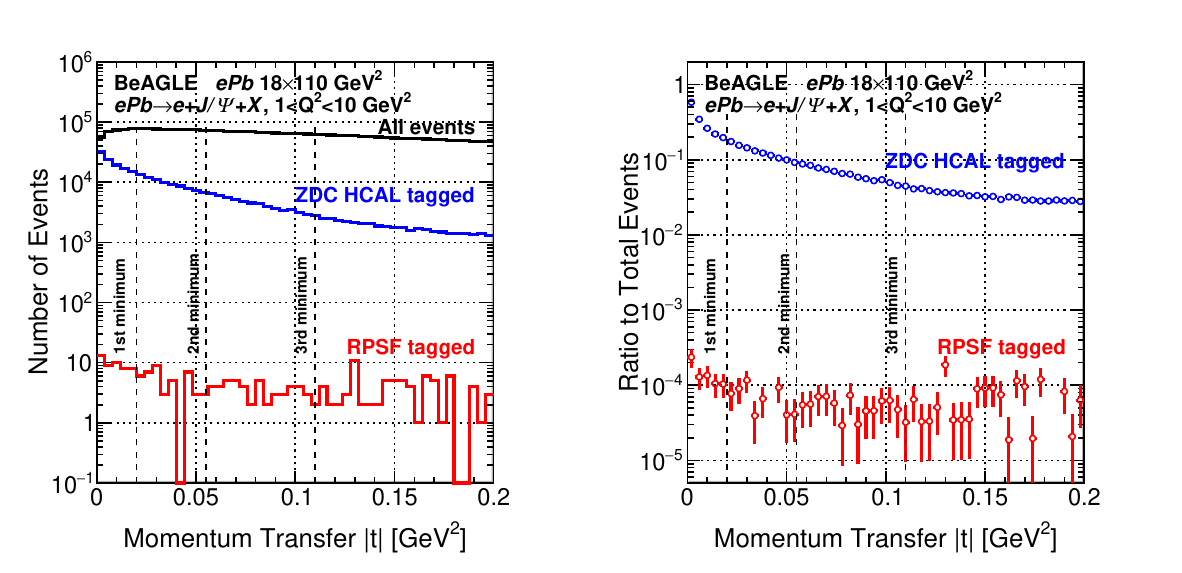}
    \caption{Left: Number of non-vetoed incoherent diffractive events in ePb collisions as a function of momentum transfer $t$. The black curve represents all incoherent events; the blue curve shows events remaining after tagging and vetoing by the Zero Degree Calorimeter (ZDC); the red curve corresponds to events surviving after combined tagging and vetoing by both the ZDC and the Roman Pot (RP). Right: Vetoing efficiency as a function of $t$, where each curve depicts the inefficiency histogram of a specific veto selection normalized to the total incoherent event sample. Only veto selections with significant impact are shown. The term “RPSF” denotes the Roman Pot located at the Secondary Focus.
Adapted from~\cite{PhysRevD.111.072013}.}
    \label{fig:vetoing_power}
\end{figure*}

By placing silicon detectors around the secondary focus, these detectors can approach closer to the beam core, substantially enhancing forward acceptance for scattered protons, ions from diffractive interactions, and nuclear fragments that would otherwise remain undetected due to their proximity to, or overlap with, the beam envelope. This configuration increases sensitivity to particles with low transverse momentum ($p_T < 200$~MeV) and those with small longitudinal momentum loss (high $x_{L} \sim \frac{p_{\textrm{proton}}}{p_{\textrm{beam}}}$), which corresponds to minimal magnetic rigidity loss and trajectories very close to the nominal beam orbit.

Importantly, improved detection of nuclear fragments facilitates better separation of coherent and incoherent diffractive events—a critical requirement for 3D imaging of nuclei. It may also enable partial or full momentum reconstruction of the detected fragments, further advancing the EIC nuclear physics program. The left panel of Fig.~\ref{fig:vetoing_power} shows the number of non-vetoed incoherent diffractive events in ePb collisions as a function of momentum transfer $t$, while the right panel displays the ratio of non-vetoed to total incoherent events. These plots demonstrate the tagging performance of the Zero Degree Calorimeter (ZDC) and the Roman Pot at the Secondary Focus (RPSF) as functions of $t$. Notably, combining ZDC and RPSF tagging yields a significant improvement in veto efficiency. Unlike the IR-6 design, where Roman Pot tagging of heavy nuclear fragments is limited, the IR-8 design offers much more effective tagging due to the smaller beam size at the secondary focus—comparable to that at the interaction point—allowing the Roman Pot detectors to be positioned closer to the beam core.

\section{Summary}
As discussed, the design of the second detector and its interaction region facilitates cross-checking and cross-calibration with the first detector, providing valuable opportunities to minimize systematic uncertainties. Beyond complementarity, the second detector can significantly support and enhance the overall EIC science program by introducing novel features in the second interaction region design. These enhancements are particularly impactful for the exclusive, tagging, and diffractive physics programs, thereby expanding the EIC’s scientific reach and discovery potential.

\end{document}